\documentclass[aps,prc,a4paper,twocolumn, noshowpacs, superscriptaddress,floatfix]{revtex4}
\usepackage{CJK}
\usepackage{graphicx}
\usepackage{subfigure}
\usepackage{color}
\usepackage{amsmath}
\usepackage{amssymb}

\begin{document}
    \title{ \Large Quantum coherence and entanglement under the influence of decoherence}
      
     \author{Wen-Yang Sun}   
    \email[Corresponding author: ]{swy$_$3299@163.com (W.-Y. Sun)}
    \affiliation{School of Electrical  and Electronic Engineering, Anhui Science and Technology University, Bengbu 233030, People's Republic of China}
    \affiliation{School of Physics and Optoelectronics Engineering, Anhui University, Hefei 230601, People's Republic of China}
    \affiliation{Key Laboratory of Functional Materials  and Devices for Informatics of Anhui Higher Education Institutes, Fuyang Normal University, Fuyang 236037, People's Republic of China}
    
    \author{A-Min Ding}  
    \affiliation{School of Electrical  and Electronic Engineering, Anhui Science and Technology University, Bengbu 233030, People's Republic of China}
    
    \author{Juan He} 
    \affiliation{Key Laboratory of Functional Materials  and Devices for Informatics of Anhui Higher Education Institutes, Fuyang Normal University, Fuyang 236037, People's Republic of China}
    
    \author{Jiadong Shi} 
   \affiliation{Key Laboratory of Functional Materials  and Devices for Informatics of Anhui Higher Education Institutes, Fuyang Normal University, Fuyang 236037, People's Republic of China}
    \affiliation{School of Computer and Information Technology, Anhui Vocational and Technical College, Hefei 230011, People's Republic of China}
    
    \author{Le Wang} 
    \affiliation{School of Electrical  and Electronic Engineering, Anhui Science and Technology University, Bengbu 233030, People's Republic of China}
    
     \author{Hui-Fang Xu} 
    \affiliation{School of Electrical  and Electronic Engineering, Anhui Science and Technology University, Bengbu 233030, People's Republic of China}
    
    \author{Dong Wang}  
    \affiliation{School of Physics and Optoelectronics Engineering, Anhui University, Hefei 230601, People's Republic of China}
    
    \author{Liu Ye}  
    \email[Corresponding author: ]{yeliu@ahu.edu.cn (L. Ye)} 
    \affiliation{School of Physics  and Optoelectronics Engineering, Anhui University, Hefei 230601, People's Republic of China}
    
 \vspace{10pt}
 
	\begin{abstract}
 {\normalsize	\textbf{Abstract:} In this work, we delve into the dynamic traits of the relative entropy of quantum coherence (REQC) as the quantum system interacts with the different noisy channels, drawing comparisons with entanglement (concurrence). The research results demonstrate  the broader prevalence and stronger robustness of the REQC as opposed to concurrence. It's worth noting that  the bit flip channel cannot  uphold a constant nonzero frozen the REQC, besides, the concurrence follows a pattern of temporary reduction to zero, followed by recovery after a certain time span. More importantly, the REQC maintains its presence consistently until reaching a critical threshold, whereas concurrence experiences completely attenuation to zero under the influence of phase damping and amplitude damping channels.
 	
	}
	
	\end{abstract}
    \maketitle

   \section{Introduction} \label{sec1}
   
   Quantum coherence, stemming from the principles of quantum superposition, holds a crucial significance in the field of physics \cite{w1,w2,w3,w4,w5,w6,w7,w8,w9,w10,w11}. It stands as a common prerequisite for both entanglement and various other forms of quantum correlations. In addition, it also functions as a valuable asset in the domain of quantum metrology  \cite{w12,w13}, facilitating the execution of specific quantum computation and quantum information processing (QIP) operations, including tasks like quantum teleportation and encryption  \cite{w14,w15,w16,w17,w18,w19,w20}. To date, the precise quantification of quantum coherence has remained a prominent topic within the realm of quantum information and computation, with the recent introduction of a rigorous framework for its quantification. Notably, two quantification measures based on the   norm and relative entropy have been successfully applied to assess quantum coherence  \cite{w21}.
   
   On the other hand, quantum entanglement \cite{w22}, assessed using the Wootters' concurrence \cite{w23}, embodies a robust nonclassical correlation inherent in bipartite quantum systems. This exceptional attribute designates it as a fundamental resource for a multitude of the QIP tasks. However, in almost all real-world quantum information processing task scenarios, quantum decoherence \cite{w24,w25,w26,w27,w28,w29,w30,w31,w32,w33} emerges as a consequence of the unavoidable interaction between a quantum system and its external noise. This interaction results in the degradation of quantum entanglement, and leads to a phenomenon known as entanglement sudden death \cite{w34,w35} under specific conditions. Even though we acknowledge the connection between quantum coherence and entanglement, the precise nature of their interrelation remains enigmatic. This knowledge gap has ignited our profound curiosity in investigating the intricate relationship between these two quantum measures and their dynamic responses when subjected to decoherence channels. 
   
   Consequently, in this work, we center our attention on the relative entropy of quantum coherence (REQC) in comparison to concurrence, examining their dynamic behaviors characteristics under the context of the different noisy environments.
   
   This letter is organized as follows. In Sec. \ref{sec2} we briefly review the REQC and one entanglement witness, concurrence. In Sec. \ref{sec3} we study the evolution of the REQC and concurrence when the system is subjected to different decoherence channels. Finally, we end up our letter with a brief conclusion.
   
   \section{A briefly review for two quantum measures} \label{sec2}
   
   \begin{figure}
   	\centering
   	\includegraphics[scale=0.33]{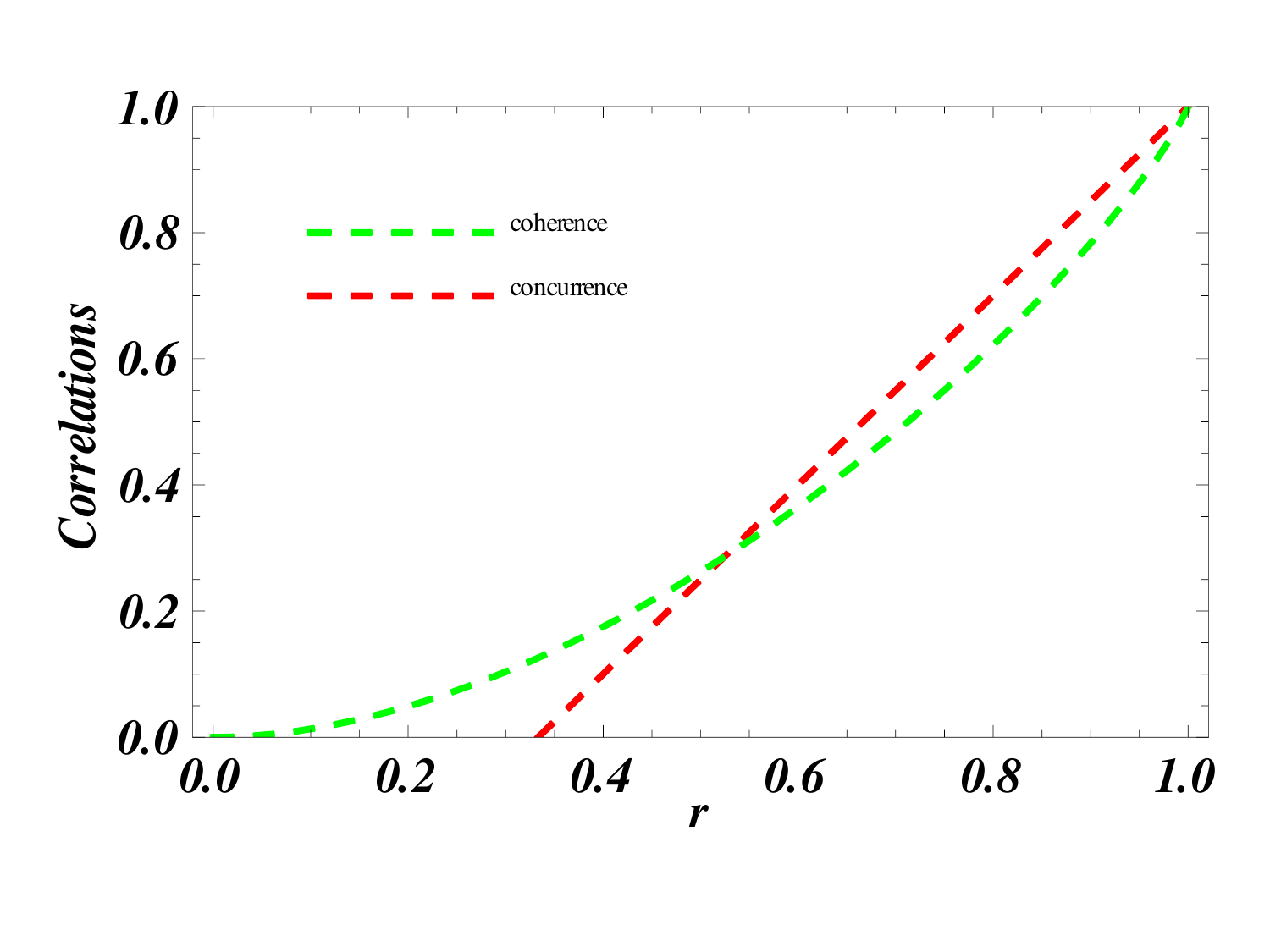}
   	\caption{\small  (Color online) The coherence (REQC) and concurrence versus the SP $r$.}
   	\label{fig1}
   \end{figure}

   Let us commence by providing a concise overview of the definitions and formalisms pertaining to two quantum measures, specifically, quantum coherence (REQC) and entanglement (concurrence). Firstly, for any quantum state $\rho$ within the Hilbert space $H$ , quantum coherence can be quantified in a convenient manner by utilizing  the relative entropy between the state  $\rho$  and its nearest incoherent state $\sigma $, denoted as \cite{w21, w30}
   \begin{equation}\label{E1} 
   \begin{split}  
  {C_{re}}\left( \rho  \right): = \mathop {\min }\limits_{ \sigma  \in \mathcal{I}} S\left( {\rho \parallel \sigma } \right),  
     \end{split}  
   \end{equation}
   where $ S(\rho \parallel \sigma ) = Tr(\rho {\log _2}\rho  - \rho {\log _2}\sigma )$. In particular, there exists a closed-form solution that simplifies the assessment of analytical expressions, as provided by the following
   \begin{equation}\label{E2} 
   \begin{split}  
    {C_{re}}\left( \rho  \right) = S\left( {{\rho _d}} \right) - S\left( \rho  \right),
     \end{split}   
   \end{equation}
   where  $ S(\rho ) =  - Tr\rho \log _2 \rho $ is the von Neumann entropy, and $ {\rho _{\rm{d}}}{\rm{ = }}\sum\limits_i {{\rho _{ii}}\left| i \right\rangle } \left\langle i \right|$ is the diagonal elements of the quantum state $\rho$.
   
    \begin{figure*}
   	\begin{minipage}[t]{0.48\linewidth}
   		\centering
   		\includegraphics[scale=0.32]{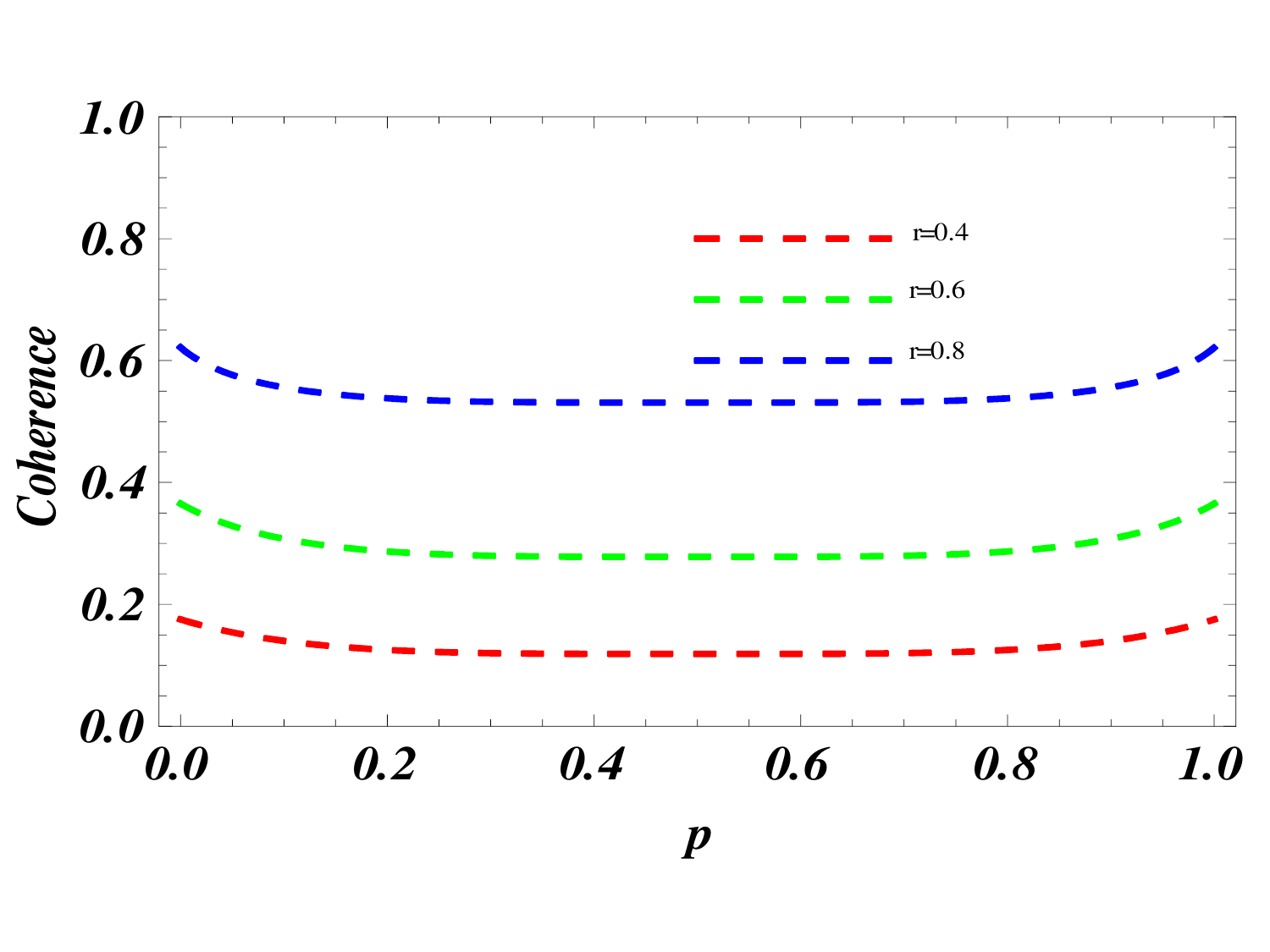}
   	\end{minipage}%
   	\begin{minipage}[t]{0.48\linewidth}
   		\centering
   		\includegraphics[scale=0.315]{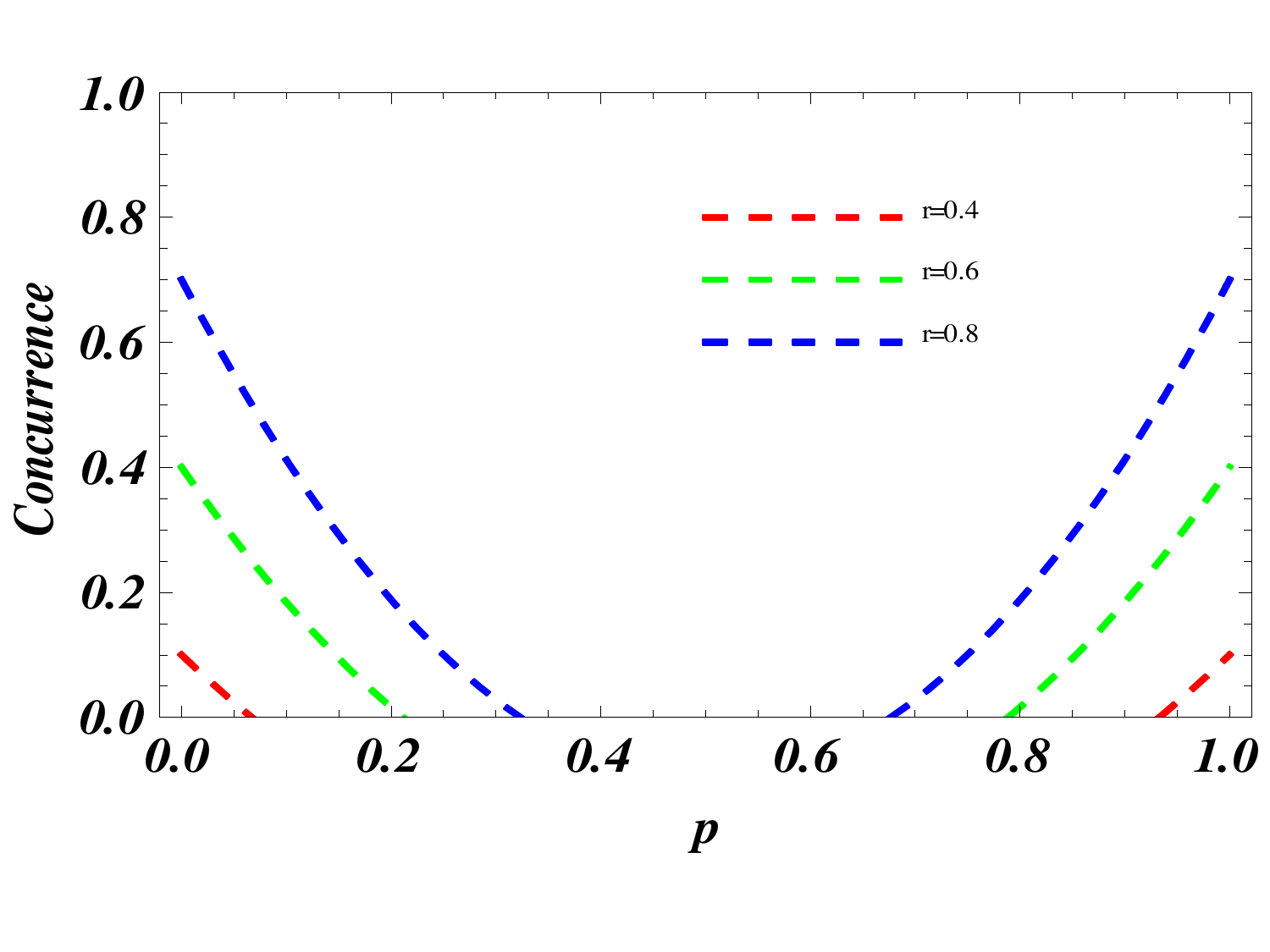}
   	\end{minipage}
   	\caption{\small  (Color online) The dynamical behaviors of coherence (REQC) (left) and concurrence (right) as a function of decoherence parameter  $p$  for the different  SP $r$  under the bit flip channel.}
   	\label{fig2}
   \end{figure*} 

\begin{figure*}
	\begin{minipage}[t]{0.48\linewidth}
		\centering
		\includegraphics[scale=0.32]{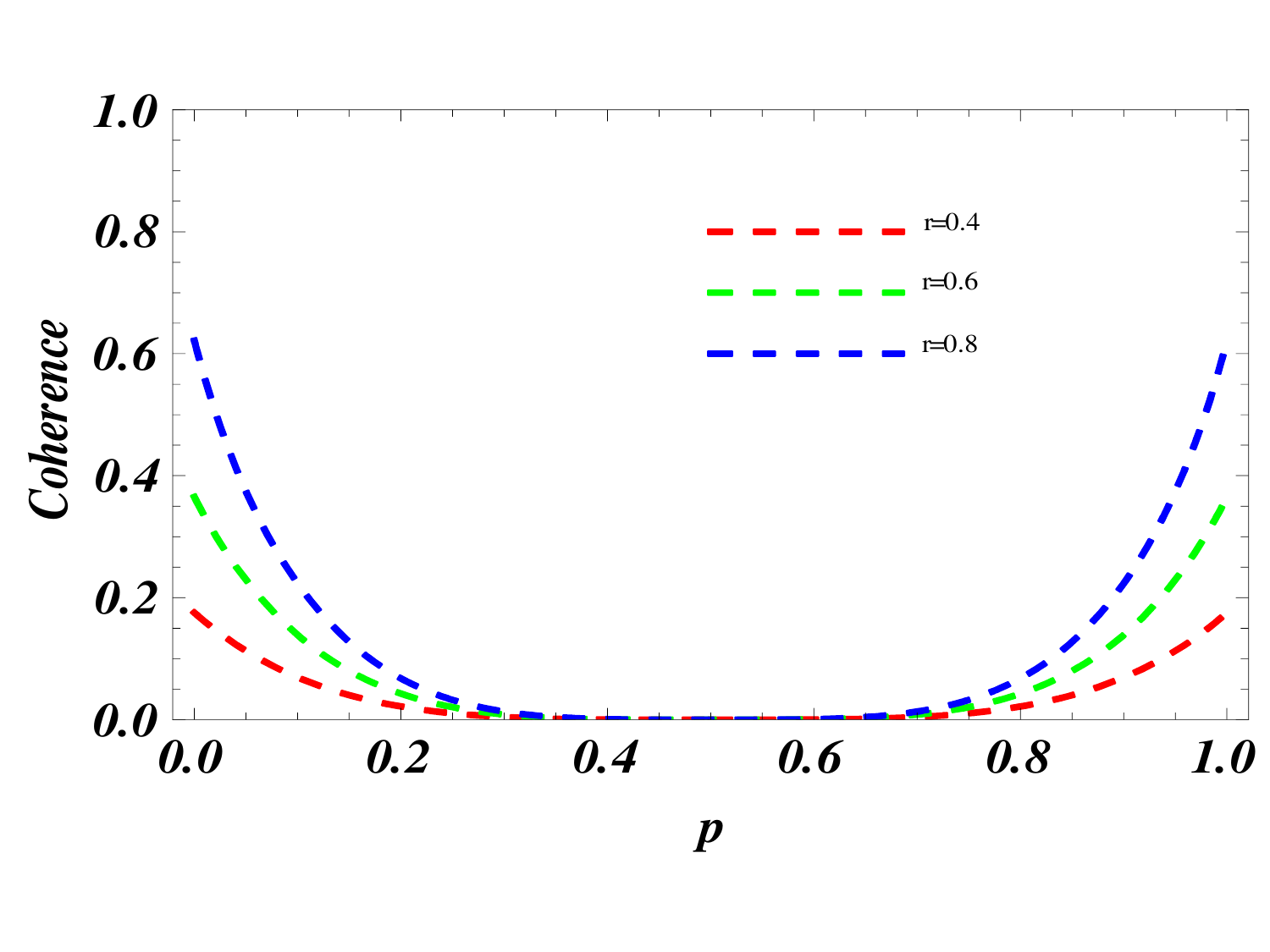}
	\end{minipage}%
	\begin{minipage}[t]{0.48\linewidth}
		\centering
		\includegraphics[scale=0.32]{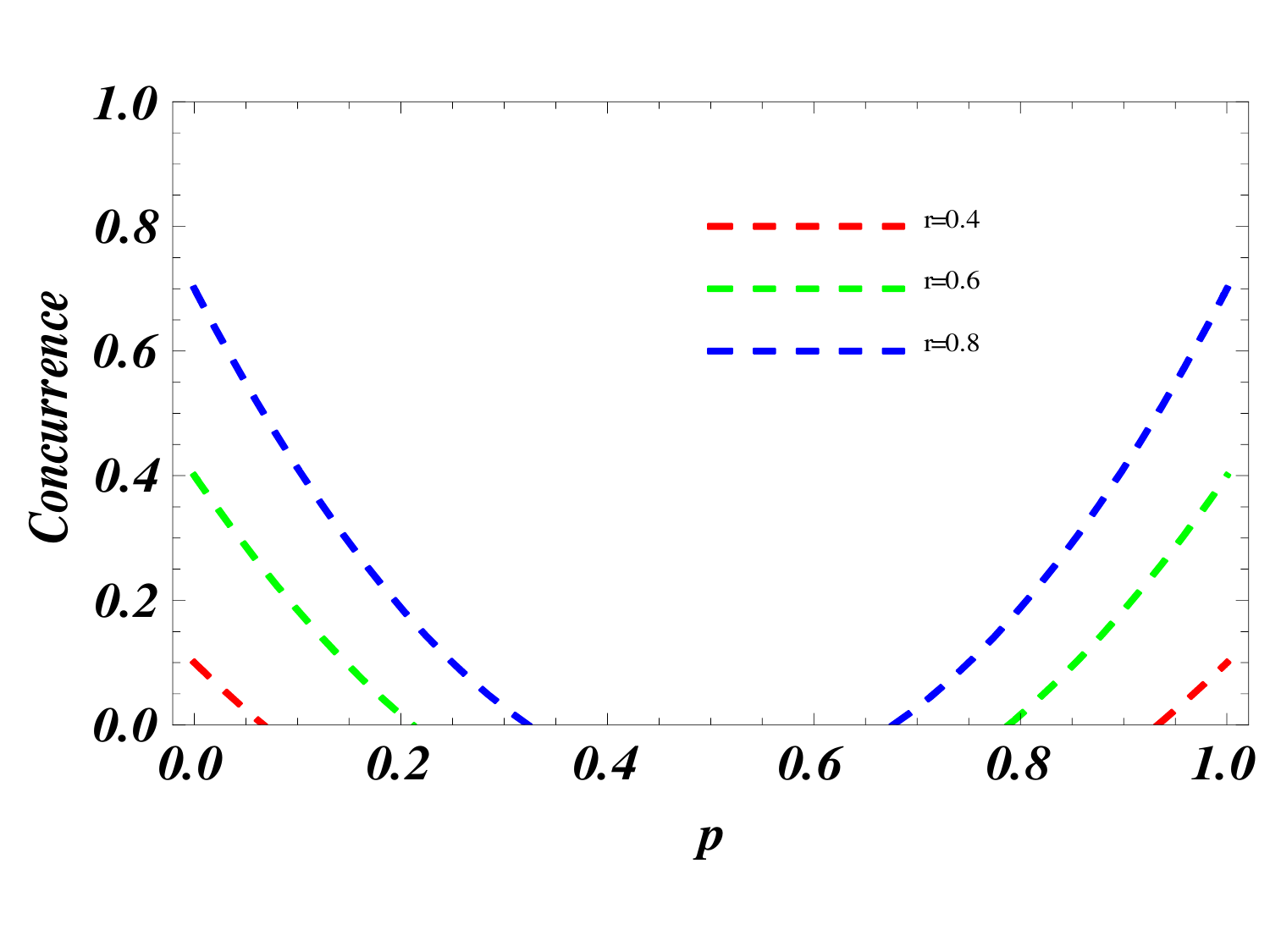}
	\end{minipage}
	\caption{\small  (Color online) The dynamical behaviors of coherence (REQC) (left) and concurrence (right) as a function of decoherence parameter  $p$ for the different SP $r$  under the phase flip channel.}
	\label{fig3}
\end{figure*}  

   Additionally, the degree of entanglement for a mixed two-qubit system can be conveniently quantified by concurrence \cite{w22,w23}. Generally speaking, concurrence can be defined as 
   \begin{equation}\label{E3} 
   \begin{split}
  C\left( \rho  \right)= \max \left\{ {0,\sqrt {{\varepsilon _1}}  - \sqrt {{\varepsilon _2}}  - \sqrt {{\varepsilon _3}}  - \sqrt {{\varepsilon _4}} } \right\},
   \end{split}   
   \end{equation}
   where ${\varepsilon _1} \ge {\varepsilon _2} \ge {\varepsilon _3} \ge {\varepsilon _4} \ge 0$, and ${\varepsilon _i}(i = 1,{\rm{ }}2,{\rm{ }}3,{\rm{ }}4)$ are the eigenvalues of the matrix     $ R = \rho \left( {\sigma _1^y \otimes \sigma _2^y} \right){\rho ^ * }\left( {\sigma _1^y \otimes \sigma _2^y} \right)$. Fortunately, for the two-qubit X-state ${\rho_X}$, concurrence can be expressed more succinctly as shown in following
   \begin{equation}\label{E4} 
   \begin{split}
     C\left({\rho _X}  \right) =2 \max\left\{ {0,|\rho_{23}| - \sqrt {{\rho _{11}}{\rho _{44}}} } \right., 
   {\rm{            }}\left.{|\rho_{14}|  - \sqrt {{\rho _{22}}{\rho _{33}}} } \right\}.
    \end{split}   
   \end{equation}
   where  ${\rho _{ij}}$ are the elements of the matrix  ${\rho _X}$. Subsequently, we will investigate the dynamic characteristics of the REQC and concurrence when the initial system interacts with its exterior noises.
   
   \section{Dynamics behaviors of the REQC and concurrence under decoherence noises} \label{sec3}
   
   In this section, we focus on the interaction between the Werner state and its external noise, and we conduct an in-depth examination of the dynamic properties of both REQC and concurrence. The specific form of the Werner state is as follows
   \begin{equation}\label{E5} 
   \begin{split}
   {\rho _{AB}} = r \left| {{\psi ^ - }} \right\rangle \left\langle {\psi {}^ - } \right| + \frac{{1 - r}}{4}I \otimes I,
   \end{split}   
   \end{equation}
   where  $\left| {{\psi ^ - }} \right\rangle  = {{\left( {\left| {01} \right\rangle  - \left| {10} \right\rangle } \right)} \mathord{\left/
   		{\vphantom {{\left( {\left| {01} \right\rangle  - \left| {10} \right\rangle } \right)} {\sqrt 2 }}} \right.
   		\kern-\nulldelimiterspace} {\sqrt 2 }}$ represents a maximally entangled state, $I$ represents the single-qubit identity operator, and the state parameter (SP) $r \in [0,1]$. It’s important to note that the Werner state is disentangled as the SP $r$ is less than  $1/3$, and it becomes an entangled state when the SP $r$  is larger than  $1/3$.
    	       
 	  \begin{figure*}
 	  	\begin{minipage}[t]{0.48\linewidth}
 	  		\centering
 	  		\includegraphics[scale=0.32]{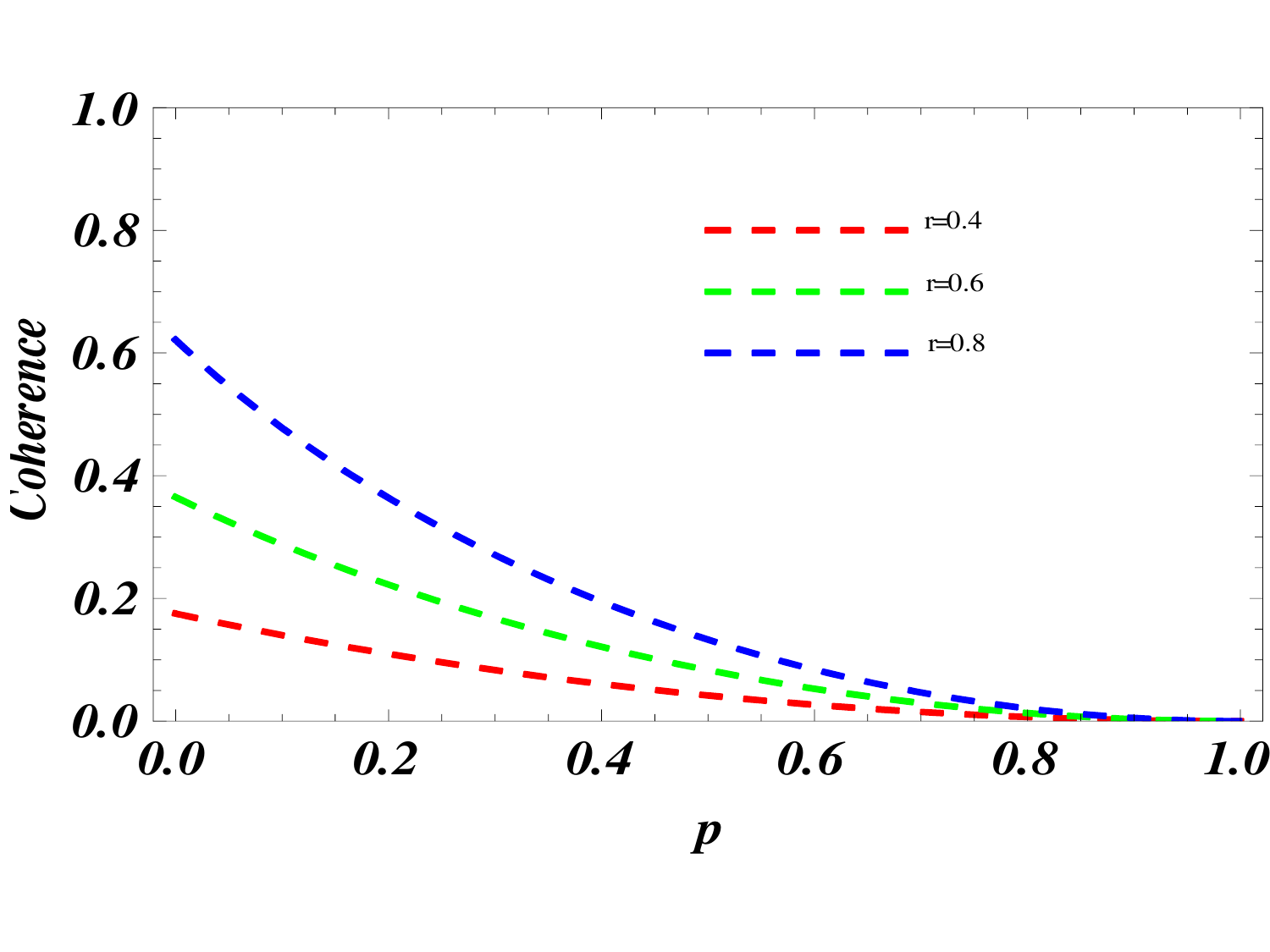}
 	  	\end{minipage}%
 	  	\begin{minipage}[t]{0.48\linewidth}
 	  		\centering
 	  		\includegraphics[scale=0.32]{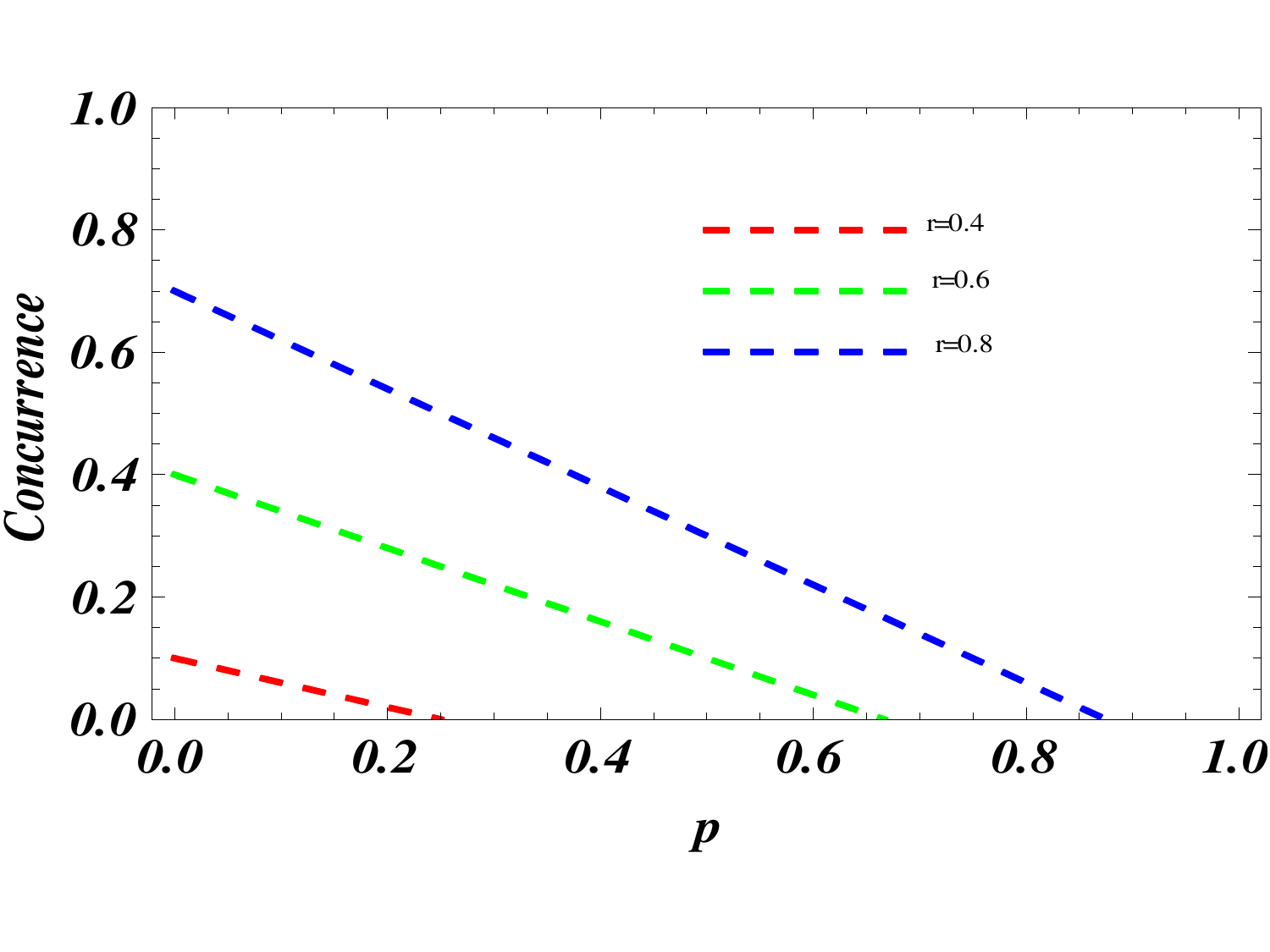}
 	  	\end{minipage}
 	  	\caption{\small  (Color online) The dynamical behaviors of coherence (REQC) (left) and concurrence (right) as a function of decoherence parameter  $p$ for the different SP $r$  under the phase damping channel.}
 	  	\label{fig4}
 	  \end{figure*}     
   
As per the equations provided in Eqs. (\ref{E1}), (\ref{E2}) and (\ref{E5}), the graphical representations of quantum coherence (REQC) and entanglement (concurrence) concerning the SP  $r$ are depicted in FIG.  \ref {fig1}. These plots reveal that the REQC always persists for arbitrary states  $0 \le r \le 1$, whereas concurrence is only present in the case of the SP  $1 \ge r > 1/3$. Intriguingly, it's also noteworthy that the REQC surpasses concurrence when the SP  $0 < r < 0.52$, but falls below it as the SP  $0.52 < r < 1$. Notably, when the  SP $r$  is equal to 0 and $1$, the REQC equals concurrence. To sum up, our findings affirm that quantum coherence exhibits a broader prevalence and stronger robustness compared to quantum entanglement.

Subsequently, we will explore how decoherence channels affect quantum correlations when Werner states interact with local decoherence channels. In this context, we analyze the interaction between the quantum system and its exterior noisy environment using the operator-sum representation formalism. Taking advantage of the Kraus operators' approach, the evolved state under the influence of a local noise can be described as a trace-preserving quantum operation, as represented in following
\begin{equation}\label{E6} 
\begin{split}
{\rho _{AB}}\left( t \right) = \sum\limits_{i = 0,j = 0}^1 {\left( {K_{_i}^A \otimes K_{_j}^B} \right)} \rho {\left( {K_{_i}^A \otimes K_{_j}^B} \right)^\dag },
\end{split}   
\end{equation} 
where  ${K_{i,j}}$ are the Kraus operators, meeting the trace-preserving condition  $\sum\nolimits_{i,j} {K_{i,j}^\dag } {K_{i,j}} = I$. In the subsequent discussion, we will examine the dynamic characteristics  of the REQC and concurrence under the flip, phase damping, and amplitude damping channels.

\subsection{Flip channel} \label{sec3.1}

In practical, local noisy channels can be effectively represented by several common quantum errors, including bit flip, bit-phase flip, and phase flip errors. Correspondingly, the Kraus operators for each of these errors are defined as follows
\begin{equation}\label{E7} 
\begin{split}
{K_0} = \sqrt {1 - p} {\rm{ }}I,{\rm{      }}{K_1} = \sqrt p {\sigma _i},
\end{split}   
\end{equation} 
here, ${\sigma _i}$  represents the Pauli operators, with $i = x$  represents the bit flip channel,  $i = y$ corresponds to the bit-phase flip channel, and $i = z$  denotes the phase flip channel. These sets can be readily understood as the probability $1-p$  of staying in the same state and the probability $p$ of encountering an error.

Firstly, let us examine the impact of the bit flip on the Werner state, and then we show the dynamic behaviors of the REQC and concurrence as functions of the decoherence parameter $p$ as the value of SP  $r$ is different   within FIG. \ref {fig2}. As shown in  FIG.  \ref {fig2}, in the bit flip channel, we observe that as the decoherence parameter $p$ increases, the REQC experiences a smooth transition, however, it cannot be frozen after a certain period. In addition,  concurrence is rapidly lost entirely, resulting in a phenomenon commonly known as entanglement sudden death. Remarkably, after a dark death interval, concurrence begins to recover. This leads us to the conclusion that coherence is generally more resilient and robust to decoherence compared to entanglement (concurrence) under the bit flip channel. Additionally, it's worth noting that the bit-phase flip channel has a similar impact on both coherence and concurrence as the bit flip channel.

\begin{figure*}
	\begin{minipage}[t]{0.45\linewidth}
		\centering
		\includegraphics[scale=0.315]{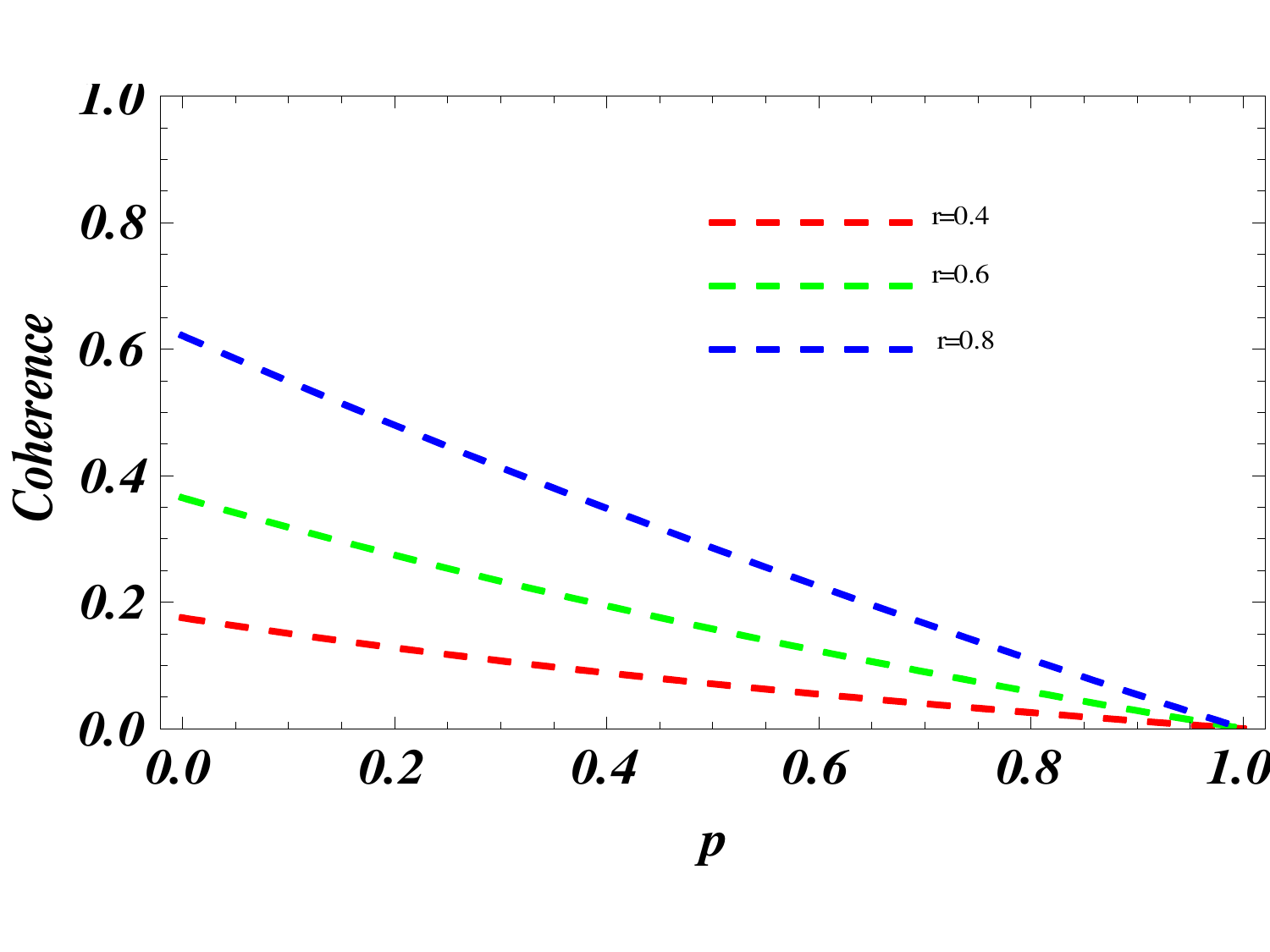}
	\end{minipage}%
	\begin{minipage}[t]{0.45\linewidth}
		\centering
		\includegraphics[scale=0.32]{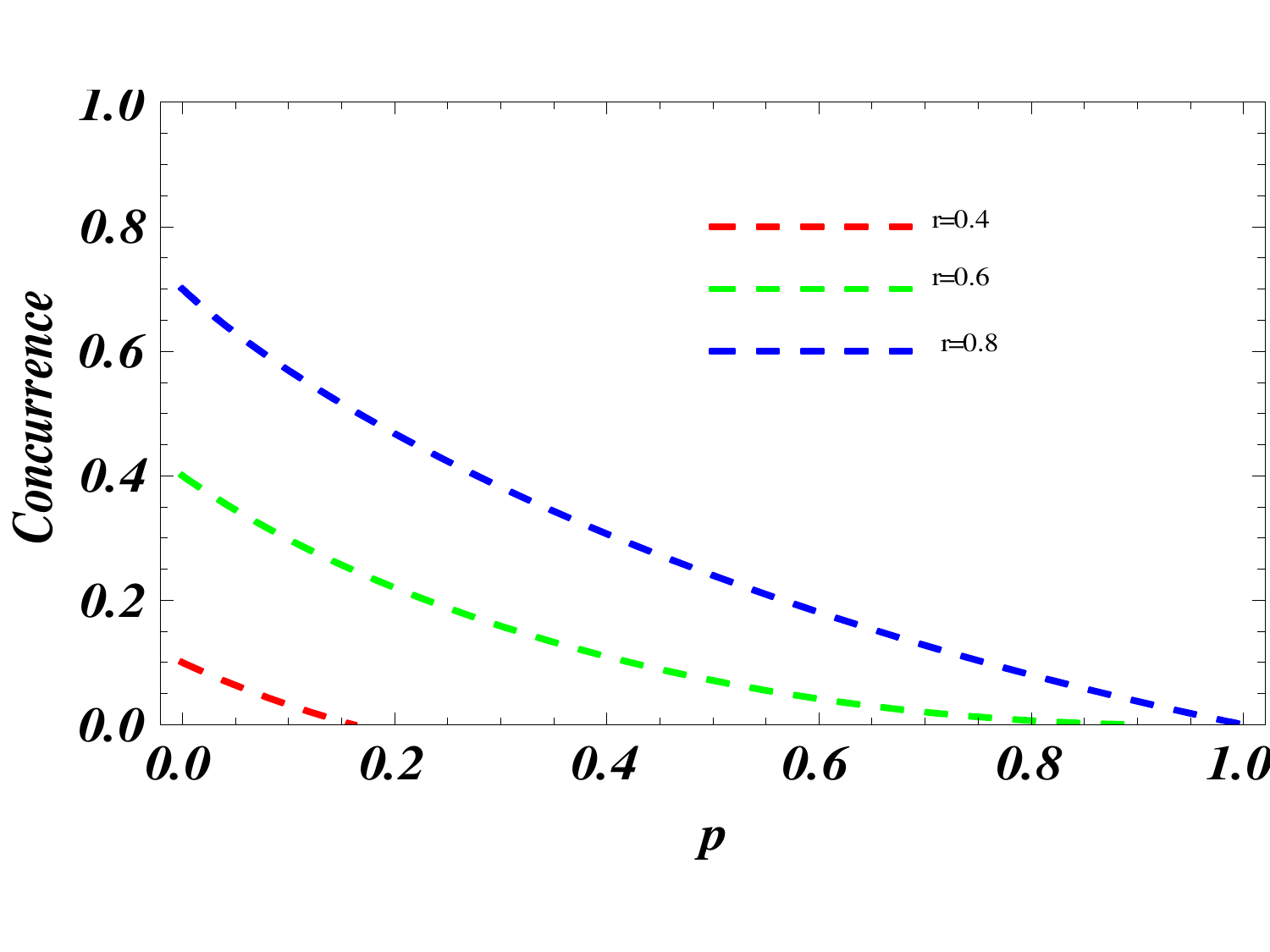}
	\end{minipage}
	\caption{\small  (Color online) The dynamical behaviors of coherence (REQC) (left) and concurrence (right) as a function of decoherence parameter  $p$ for the different initial states under the amplitude damping channel.}
	\label{fig5}
\end{figure*} 

Next, we delve into the effects of the phase flip channel in detail and present the dynamic behaviors of the REQC and concurrence concerning the decoherence parameter  $p$  in FIG. \ref{fig3}. From the figure, it's apparent that both the REQC and concurrence grow rapidly with the increase of SP  $r$ , when the decoherence parameter is a larger value or a smaller value, such as equal to $0.1$ or $0.9$.  Astonishingly, as the SP $r$ is a certain value, due to the presence of decoherence, the REQC suddenly disappears but then can reappear and continue to increase to the initial value. Similar to the REQC, concurrence is at times completely lost but subsequently shows signs of recovery.

\subsection{Phase damping channel} \label{sec3.2}

Now, let us delve into the dynamic characteristics of both REQC and concurrence under the phase damping channel. This channel can be described as the loss of quantum coherence without a loss of energy, and the corresponding Kraus operators can be given by
\begin{equation}\label{E8} 
\begin{split}
{K_0} = \left( {\begin{array}{*{20}{c}}
	1&0\\
	0&{\sqrt {1 - p} }
	\end{array}} \right),{\rm{               }}{K_1} = \left( {\begin{array}{*{20}{c}}
	0&0\\
	0&{\sqrt p }
	\end{array}} \right),
\end{split}   
\end{equation} 
where the relation between the decoherence parameter $p$  and time $t$  is expressed as $p = 1 - {e^{ - \gamma t}}$ with $\gamma $ is the decay rate. Later, based on the formulas provided in Eqs. (\ref{E2}), (\ref{E3}), (\ref{E5}) and (\ref{E8}), the graphical representations of the REQC and concurrence concerning the decoherence parameter  $p$ are presented in FIG. \ref{fig4}.

As depicted in FIG. \ref{fig4}, we can discern that the REQC remains non-zero for all values of the decoherence parameter  $p$ except at the specific point  $p=1$. However, the concurrence fleetly diminishes to zero with the increase of the decoherence parameter $p$. Moreover, at the specific case, the SP is equal to $1$, the concurrence is always non-zero except at $p=1$. This observation highlights the greater robustness of the REQC compared to concurrence. It suggests that the REQC can offer more insights into the characteristics of the quantum system under consideration. Consequently, we hold the opinion that quantum coherence is generally more robustness to decoherence compared to entanglement.

\subsection{Amplitude damping channel} \label{sec3.3}

The amplitude damping channel characterizes a classical noise process that accounts for the dissipative interaction between quantum system and its surrounding environment. One simple approach to understanding this process is by examining the corresponding Kraus operators, which are defined as follows
\begin{equation}\label{E9} 
\begin{split}
{K_0} = \left( {\begin{array}{*{20}{c}}
	1&0\\
	0&{\sqrt {1 - p} }
	\end{array}} \right),{\rm{                 }}{K_1} = \left( {\begin{array}{*{20}{c}}
	0&{\sqrt p }\\
	0&0
	\end{array}} \right),
\end{split}   
\end{equation} 
where $p$  signifies oscillations that illustrate how the decay of the atom's excited state is influenced by the coherent interactions between the quantum system and the external noisy environment.

Then, we present the graphical representations of both REQC and concurrence as functions of the decoherence parameter  $p$ in FIG. \ref{fig5}. As shown in FIG. \ref{fig5}, we clearly acquire that the amplitude damping channel has a similar impact on both REQC and concurrence as the phase damping channel. Thus, we will not reiterate the results here.

\section{Conclusions}\label{sec4}

In this letter, we have explored the dynamic  characteristics of quantum coherence quantified by the relative entropy in comparison to concurrence under the influence of the different noisy channels. Our findings reveal that quantum coherence is a more ubiquitous manifestation than entanglement, although there are some scenarios where concurrence may exceed the REQC for specific states. Notably, we have demonstrated that quantum coherence generally exhibits greater robustness against decoherence than that entanglement. Besides, we have observed that the bit and bit-phase flip channels  cannot permit nonzero frozen coherence.  Additionally, entanglement eventually diminishes to zero and then revives after a certain period. Furthermore, when a particular initial state is subjected to phase and amplitude damping channels, both REQC and concurrence decrease steadily with the increase of the decoherence parameter   $p$. Significantly, the notable distinction between them is that the REQC persists except at the critical point   $p=1$, whereas concurrence inevitably diminishes to zero.

\section*{Acknowledgement} 
This work was supported by the Anhui Provincial Natural Science Foundation under the Grant No. 2008085QF328, and the Natural Science Foundation of Education Department of Anhui Province under Grant Nos. 2023AH051859, 2022AH040235, 2022AH051634, KJ2021ZD0071, 2023AH051845 and KJ2021A0867, and the National Science Foundation of China under Grant Nos. 12205047 and 11847020, and the Open Project Program of Key Laboratory of Functional Materials and Devices for Informatics of Anhui Higher Education Institutes (Fuyang Normal University) under Grant No. FSKFKT003, and the Talent Introduction Project of Anhui Science and Technology University under Grant Nos. DQYJ202005 and DQYJ202004. And also, the Horizontal Research Project of the Design of 10-bit 8-Channel ADC Module Based on SMIC 40nm LL (Grant No. 880937).

    \end{document}